\documentclass{llncs}

\usepackage[pdftex]{graphicx}
\usepackage{epstopdf}        
     
\usepackage{makeidx}  
\usepackage{amsmath}
\usepackage{amsfonts,amssymb}
\usepackage{clrscode3e}        
\usepackage{multirow}

\newcommand{\dt}{\displaystyle}

\newcommand{\mbf}[1]{\protect\text{\boldmath$#1$}}                 

\begin{document}
\frontmatter          
\pagestyle{headings}  
\addtocmark{The set cover problem} 
\title{The interval greedy algorithm\\ for discrete optimization problems\\ with interval objective function}
\titlerunning{The interval greedy algorithm}  
%
\author{Alexander Prolubnikov}
\authorrunning{Alexander Prolubnikov} 
%
\tocauthor{Alexander Prolubnikov}
\institute{Omsk State University, Omsk, Russian Federation\\
\email{a.v.prolubnikov@mail.ru}
}
\maketitle              

\begin{abstract}

We consider a wide class of the discrete optimization problems with interval objective function. We give a generalization of the greedy algorithm for the problems. Using the algorithm, we obtain the set of all possible greedy solutions and the set of all possible values of~the~objective function for the solutions. For~a~given probability distribution on intervals of objective function' coefficients, we compute probabilities of the solutions, compute expected values of~the~objective function for them and other probabilistic characteristics of the problem.

\end{abstract}

\section*{Introduction}

A great variety of applied problems may be formulated as discrete optimization problems. There may be uncertainties in input data for an applied problem and the discrete optimization methods that operates with exact values of weights will not give us any more than information on some of many possible solutions which correspond to some possible values of the input data. It is not always a reliable way to use the mean values of inexact parameters since they may be unrepresentative. For different possible values of inexact parameters, there may be different optimal solutions with different values of objective function. And the difference may be big enough. 

The uncertainties on the input data may be caused by various reasons. It may be measurements errors. It may be the case that the values of some parameters are varied. Thus, for example, the amount of fuel that is needed to take the same load to the same point by vehicle is different for different weather conditions and  different fuel quality.

It is often the case that an interval of possible values is the only known information on uncertain parameter of an optimization problem. Sometimes we may have an information on probability distribution of the parameter's values on the interval. 

Optimization problems with inexact input data has been investigated in many directions by many researchers. Linear programming problems with inexact input data have been considered in \cite{Falk,Vatolin,AgayanRyutinTikhonov,Hladik,Fiedler} in particular. As usual, the presented approaches search for some unique solution of the problem. The robust optimization \cite{KasperskyZielinski,KasperskyZielinski2,Kaspersky,BertsimasSim,KouvelisYu,YuKouvelis2} is an example of a such approach. Other approaches search for some predefined set of the possible solutions that corresponds to some possible values of inexact parameters. They are presented in \cite{YamanKarasanPinar,KozinaPerepelitsa,PerepelitsaTebueva,Kozina,Demchenko}. 

We consider the discrete optmization problems with interval objective function coefficients. Using the approach we present, in the situation of uncertainty the person that make a decision may obtain possible approximate solutions, the possible values of objective function for them and other information that may be used to analyze the possible scenarios for the situation. We give a generalization of the greedy algorithm for the case of interval objective function. This algorithm gives a set of solutions for possible values of the function's coefficients. Such a set consists of exact or approximate solutions for its possible values. Also, it gives a set of possible values of objective function for the solutions. For a given probability distribution on intervals of function' coefficients, we compute probabilities of the solutions, expected values of the objective function for them, etc.

\section{Discrete optimization problems with interval weights}

We consider the discrete optimization problems that may be formulated in the following way. Let $E\!=\!\{e_1,\ldots,e_n\}$. Let $w(e)\!>\!0$ be the weight of $e\!\in\!E$, $w_i\!=\!w(e_i)$. A binary vector $x\!=\!(x_1,\ldots,x_n)$ determines the set $E_x\!\subset\!E$: $e_i\!\in\! E_x$ iff $x_i\!=\!1$. The set of feasible solutions $\mathcal{D}$ is given. The set $\mathcal{D}$ may be considered as a set of vectors which, for some graph or hypergraph, may be associated with its subgraphs of some predefined specific form, i.e., set of the paths that connects the two graph's vertices, set of spanning trees, set of hamiltonian cycles, etc.

\bigskip

\noindent{\bf Optimization problem (I).} We need to find such $x\!\in\!\mathcal{D}$ that gives minimum of the objective function 
\begin{equation}\label{1}  
f(x,w)=\sum\limits_{e\in E_x}w(e).
\end{equation}

\bigskip

For example, if the set $E$ is the set of a graph edges, the following optimization problems on graphs may be stated in a such way: the shortest path problem, the minimum spanning tree problem, the traveling salesman problem, the minimum edge cover problem. If the set $E$ is a set of graph's vertices, we may state in a such way the minimum vertex cover problem and others. If $E$ is a set of edges of a hypergraph, in particular, we may state in a such way the set cover problem that we shall consider further. 

Not only optimization problems on graphs and hypergraphs may be formulated as the problems of the form (I). For example, the boolean knapsack problem may be formulated this way too.

\smallskip

The discrete optimization problems on graphs and hypergraphs of the form (I) with interval weights have been considered in \cite{YamanKarasanPinar,YamanKarasanPinar2,KozinaPerepelitsa,PerepelitsaTebueva,Demchenko} et al. 

Let $\mathbb{IR}$ denotes the set of intervals on $\mathbb{R}$. We shall denote interval values using bold font. For an interval $\mbf{a}\!\in\!\mathbb{IR}$, its lower and upper bounds are denoted as $\underline{\mbf{a}}$ and $\overline{\mbf{a}}$ respectively: $\mbf{a}\!=\![\underline{\mbf{a}}, \overline{\mbf{a}}]$. If $\underline{\mbf{a}}\!=\!\overline{\mbf{a}}$ then interval $\mbf{a}$ called {\it degenerated}. The sum of intervals $\mbf{a}$ and $\mbf{b}$ is defined as follows: $\mbf{a}+\mbf{b}=[\underline{\mbf{a}}+\underline{\mbf{b}},\overline{\mbf{a}}+\overline{\mbf{b}}]$. The result of multiplication of an interval by $\alpha\!\in\!\mathbb{R}_+$ is the interval $\alpha\mbf{a}=[\alpha\underline{\mbf{a}},\alpha\overline{\mbf{a}}]$. Let $\mathbb{IR}^n$ denotes the set of interval vectors of dimensions $n$. 

We consider the discrete optimization prob\-lems with interval objective functions of the form 
\begin{equation}\label{2}  
f(x)=\sum\limits_{e\in E_x}\mbf{w}(e),
\end{equation}
where the values of weights are intervals $\mbf{w}(e)$. So, we consider all of possible weights $w(e)\!\in\!\mbf{w}(e)=[\underline{\mbf{w}}(e),\overline{\mbf{w}}(e)]\!\subset\!\mathbb{R}$ of element $e$. Let $\mbf{w}=(\mbf{w}_1,\ldots,\mbf{w}_n)\!\in\!\mathbb{IR}^n$, where $\mbf{w}_i\!=\!\mbf{w}(e_i)$, $\underline{\mbf{w}}_i>0$. 

To state the formulation of the discrete optimization problem with interval objective function, we need to define a concept of an optimal solution for the problem. One of the possible ways to do this is to use the concept of the Pareto set of possible solutions \cite{KozinaPerepelitsa,PerepelitsaTebueva,Kozina,Demchenko} considering the problem as a two-criteria optimization problem, where the criteria are $$f_1(x,\mbf{w})=\sum\limits_{e\in E_x}\underline{\mbf{w}}(e)\to\min,\ f_2(x,\mbf{w})=\sum\limits_{e\in E_x}\overline{\mbf{w}}(e)\to\min.$$

The other way is to use the notions of weak and strong optimal solutions for discrete optimization problem with interval weights \cite{YamanKarasanPinar,YamanKarasanPinar2,KozinaPerepelitsa,Kozina,Prolubnikov2}. For the prob\-lem, a {\it scenario} is a vector $w\!\in\!\mbf{w}$. A scenario $w\!\in\!\mbf{w}$ sets the discrete optimization problem of the form (I) with real-valued coefficients $w$ of its objective function. A {\it weak optimal solution} (we shall call it as {\it weak solution} further) is a solution that is optimal for some scenario $w\!\in\!\mbf{w}$. For a discrete optimization problem with an interval objective function, a {\it strong optimal solution} (we shall call it as {\it strong solution} further) is the solution that is optimal for any scenario $w\!\in\!\mbf{w}$. Note that a strong solution is a weak one too. 

Using the concept of the strong solution, we may state the following formulation of the discrete optimization problem with interval objective function. 

\smallskip

\noindent{\bf Optimization problem (II).} For given interval weights $\mbf{w}$, we need to find a strong solution of the optimization problem with given set $\mathcal{D}$ of feasible solutions and the objective function of the form (\ref{2}).

\smallskip

But, since it is often the case that there is no strong solution for the problem, trying to analize a real life situation, we may search for set of weak solutions for all possible $w\!\in\!\mbf{w}$. {\it The united solution set} is a set of all weak solutions. Using the concept of the united solution set, we may state the discrete optimization problem of the following form.

\smallskip

\noindent{\bf Optimization problem (III).} We need to find a united solution set $\varXi$: $$\varXi=\bigl\{x\in\mathcal{D}\ \bigl |\ \exists w\in\mbf{w} \ \forall y\in\mathcal{D}\ \bigl ( f(x,w)\le f(y,w)\bigr )\bigr \}.$$

\smallskip

The problem (III) may be too hard computationally even for low dimensional cases. For example, this is the case when the corresponded problem of the form (I) with real-valued function is {\bf NP}-hard. So, we may try to solve the problem (III) approximately instead, searching for the {\it united approximate solution set} that contains the approximate solutions with guaranteed acuracy for all of the possible scenarios, e.g., {\it greedy solutions} which are obtained by the greedy algorithm. Solving such problems, we may go  beyond the exhaustive search on $w\!\in\!\mbf{w}$ and try to solve the problem by less costly means.

\section{Characterization of strong solutions}

If there exists a solution of the problem (II), we have the best situation that we may have dealing with optimization problem with objective function (\ref{2}). The theorem below gives a way to check wether a weak solution is a strong one.

{\it The worst scenario} for $x\!\in\!\mathcal{D}$ is a such scenario $w\!\in\!\mbf{w}$ that $w(e)\!=\!\overline{\mbf{w}}(e)$ for $e\!\in\!E_x$ while $w(e)\!=\!\underline{\mbf{w}}(e)$ for $e\!\in\!E\setminus E_x$. It was shown in \cite{YamanKarasanPinar} that, for the longest path problem, a weak solution is a strong solution if only it is an optimal solution for its worst scenario. The same result was obtained for the minimum spanning tree problem \cite{YamanKarasanPinar2}. Indeed, the same result may be obtained for any problem of the form (I) with interval weights.

\smallskip

\begin{theorem}
A weak solution is a strong solution if and only if it is an optimal solution for its worst scenario.
\end{theorem}

\smallskip

\begin{proof}
Let $x\!\in\!\mathcal{D}$ be an optimal solution for its worst scenario. For any $y\!\in\!\mathcal{D}$, we have 
$$f(x,\overline{\mbf{w}}_x)=\sum\limits_{e\in E_x\setminus E_y}\overline{\mbf{w}}(e)+\sum\limits_{e\in E_x\cap E_y}\overline{\mbf{w}}(e)\le\sum\limits_{e\in E_y\setminus E_x}\underline{\mbf{w}}(e)+\sum\limits_{e\in E_x\cap E_y}\overline{\mbf{w}}(e),$$ 
where $\overline{\mbf{w}}_x$ is the worst scenario for $x$. For arbitrary weights $w(e)\!\in\!\mbf{w}(e)$, it holds that 
$$\sum\limits_{e\in E_x\setminus E_y}\overline{\mbf{w}}(e)+\sum\limits_{e\in E_x\cap E_y}w(e)\le\sum\limits_{e\in E_y\setminus E_x}\underline{\mbf{w}}(e)+\sum\limits_{e\in E_x\cap E_y}w(e).$$ 
Since $$\sum\limits_{e\in E_x\setminus E_y}w(e)\le\sum\limits_{e\in E_x\setminus E_y}\overline{\mbf{w}}(e),\ \sum\limits_{e\in E_y\setminus E_x}\underline{\mbf{w}}(e)\le\sum\limits_{e\in E_y\setminus E_x}w(e),$$ 
for any scenario $w\!\in\!\mbf{w}$, we have 
$$f(x,w)=\sum\limits_{e\in E_x\setminus E_y}w(e)+\sum\limits_{e\in E_x\cap E_y}w(e)\le\sum\limits_{e\in E_y\setminus E_x}w(e)+\sum\limits_{e\in E_x\cap E_y}w(e)=f(y,w).$$ 
Thus $f(x,w)\!\le\!f(y,w)$ for any $w\!\in\!\mbf{w}$, i.e., $x$ is a strong optimal solution. 
\end{proof}

As it follows from the Theorem 1, we may obtain a weak solution using, for example, some branch and bound method for a fixed scenario. Having the weak solution, we may check wether it is a strong solution. 

\section{The generalization of the greedy algorithm\\ for the case of interval objective function}

\subsection{Greedy algorithms for the problem (I)} 

A rather common approach to the problems of the form (I) is to use an appropriate greedy algorithm to get an optimal or an approximate solution of the problem. Using the greedy algorithm, we obtain the solution $x\!\in\!\mathcal{D}$ taking the elements of $e\!\in\!E$ into $E_x$ one after another in accordance with the value of {\it selection function} $\varphi$ on $e\!\in\!E$.  The algorithm stops when a feasible solution $x\!\in\!\mathcal{D}$ is obtained this way.

The selection function is specifically defined for a particular problem of the form (I). The function $\varphi: E\to \mathbb{R}_+$ depends on weight of element $e\!\in\!E$ and other parameters of the problem instance which are specified by $e$. In the simpliest case, $\varphi(e)\!=\!w(e)$. For example, consider the set cover problem. Here, where an element $e_i$ is a set $S_i$, the set $E$ is a collection of sets that may be selected in the cover that the algorithm builds. The value $\varphi(S_i)\!=\!\varphi(w_i,|S_i|)\!=w_i/|S_i|$ depends on $w_i$ and on cardinality of the set $S_i$. 

The basic scheme of the greedy algorithm is the following one.

\begin{codebox}
\Procname{$\proc{The greedy algorithm for the problem (I)}$}
\li $E_x\leftarrow \varnothing$.
\li \If $E_x$ such that $x\!\in\!\mathcal{D}$,
\li \Then 
				{\bf output} $x$. 
\li \Else 
				select such $e_{\min}\!\in\! E$ that
              $\varphi(e_{\min})=\min\{\varphi(e)\ |\ e\!\in\!E \setminus E_x\}$,
\li           $E_x\leftarrow E_x\cup\{e_{\min}\}$, 
\li           Go to step $2$.
    \End
\end{codebox}

\smallskip

It was shown in \cite{Papadimitriou}, that, using the greedy algorithm, we obtain an optimal solution for the problem of the form (I) if its set $\mathcal{D}$ has a matroidal structure, or, in a more general way, if $\mathcal{D}$ is a gridoid \cite{BjornerZiegler}. The minimum spanning tree is an example of a such problem. For some of the problem of the form (I), e.g., for the set cover problem, the greedy algorithms are asymptotically best possible approximation algorithms. 

\subsection{The interval greedy algorithm for the set cover problem} 

As an example of the presented approach application, we consider the interval greedy algorithm for the set cover problem. Hereafter, we shall abbreviate it as {\it SCP}. 

In the weighted SCP, we are given set $U$, $m\!=\! |U|$. There is a collection $S$ of its subsets $S_i\!\subseteq\!S$, $S\!=\!\{S_1,\ldots, S_n\}$, such that  $\cup_{i=1}^n S_i\!=\!U$. A~collection of sets $S'\!=\!\{S_{i_1},\ldots, S_{i_k}\}$, $S_{i_j}\!\in\! S$, is called {\it a cover} of $U$ if $\cup_{j=1}^{k}S_{i_j}\!=\!U$. For $S_i\!\in\! S$, there are given weights $w_i\!=\!w\,(S_i)$, $w_i\!>\!0$. For a collection of sets $S'\!=\!\{S_{i_1},\ldots, S_{i_k}\}$, its weight $w\,(S')$ is equal to the sum of weights of the sets that belong to $S'$: $w(S')=\sum_{j=1}^k w(S_{i_j})$. We need to find an {\it optimal cover} of $U$, i.e., the cover of minimum weight. 

In the course of operating of the greedy algorithm for SCP with non-interval weights, we select the sets in the cover based on the values of their {\it relative weights} $w_i/|S_i|$ until all of the elements of $U$ are covered. We perform iterations of the following form.
	
\bigskip

\fbox{\parbox{0.9\textwidth}{

{\bf An iteration of the greedy algorithm:}\\

For an SCP instance $\mathcal{P}$, 
\begin{itemize}
\item[1)] select $S_q$ such that
$${w}_q/|S_q|=\min\biggl\{w_i/|S_i|\ \biggr |\ \biggl (S_i\in S\biggr ) 
              \mbox{ and } \biggl (S_i\not\subset\bigcup\limits_{S_j\in E_x} S_j\biggr )\biggr\};$$
\item[2)] add $S_q$ into $E_x$: $E_x\leftarrow E_x\cup \{S_q\}$; 
\item[3)] obtain the the SCP instance $\mathcal{P}'$: $U'\leftarrow U\setminus S_q$, 
$S_j'\leftarrow  S_j\setminus S_q$, $w'\leftarrow w$.
\end{itemize}

}}

\bigskip

As a result of the greedy algorithm's iteration, we include some set $S_q\!\in\! S$ into $E_x$
and make the transition from the SCP instance $\mathcal{P}$ with given $U$, $S$, $w$ to the instance
$\mathcal{P}'$ with $U'$, $S'$, $w'$. 

The solution that the greedy algorithm gives for some scenario $w\!\in\!\mbf{w}$ we call a {\it weak approximate solution} of SCP with interval weights. We consider an approximate solution as an ordered set of elements of $S$. A {\it united approximate solution set} of the problem is a such set $\widetilde{\varXi}$ of its covers that, for every scenario $w\!\in\!\mbf{w}$, there is $x\!\in\!\widetilde{\varXi}$ such that $x$ is an approximate solution that the non-interval greedy algorithm gives for the weights that the scenario $w$ specifies for the problem. 

Note that we does not need the sets $\varXi$ or $\widetilde{\varXi}$ to take the individual solutions from them for given $w\!\in\!\mbf{w}$. For some problems, the solutions may be obtained in polynomial time so it does not matter to have it beforehand. We need them in all their entirety to obtain an information that characterize the all of the posible solutions for the problem instance. 

Obtaining an ordered cover $x\!\in\!\widetilde{\varXi}$ in the course of the interval greedy algorithm operating, we also obtain the vector $\mbf{w}[x]\!\in\!\mathbb{IR}^n$. Its component $\mbf{w}_i[x]$ is a set of a such real valued weights from $\mbf{w}_i$ that the ordered cover $x$ is obtained for some scenario $w\!\in\!\mbf{w}[x]$. I.e. for all $w_i\!\in\!\mbf{w}_i[x]$, there exists such $w\!=\!(w_1,\ldots,w_i,\ldots,w_n)\!\in\!\mbf{w}[x]\!=\!(\mbf{w}_1[x],\ldots,\mbf{w}_i[x],\ldots,\mbf{w}_n[x])$ that the non-interval greedy algorithm gives $x$ for $w$. Note that not every vector $w\!\in\!\mbf{w}[x]$ is a such scenario. 

For an SCP instance $\mathcal{P}$ with interval weights, let us denote as $U_{\mathcal{P}}$ the set we need to cover. Let $S_{\mathcal{P}}$ denotes the collection of sets that we may use to build a cover for  the problem instance and let the vector $\mbf{w}_{\mathcal{P}}\!\in\!\mathbb{IR}^n$ be the vector of interval weights of  the sets. The interval vector $\mbf{w}_{\mathcal{P}}$ is a vector of all scenarios for the problem $\mathcal{P}$. For a collection of sets $S'\!=\!\{S_{i_1},\ldots, S_{i_k}\}$, its interval weight $\mbf{w}\,(S')$ is equal to the sum of interval weights of the sets that belong to $S'$: $\mbf{w}(S')=\sum_{j=1}^k \mbf{w}(S_{i_j})$.

The interval greedy algorithm takes an instance of SCP with interval weights and, using backtracking scheme, gives a united (or weak approximate) solution. We search for all possible weak (approximate) solutions and, as a result, we obtain the united (approximate) solution set $\widetilde{\varXi}$ performing iterations of the following form.

\bigskip

\fbox{\parbox{0.9\textwidth}{

\noindent {\bf An iteration of the interval greedy algorithm for SCP}\\

For the SCP instance $\mathcal{P}$.
\begin{itemize}
\item[1)] get the set $Q=\{S_{i_1},\ldots, S_{i_t}\}$ ($|Q|=t$), 
where $S_{i_j}\!\in\! Q$\\ iff $\exists w\!\in\!\mbf{w}_{\mathcal{P}}$ such that
$${w}_{i_j}/|S_{i_j}|=\min\biggl\{w_i/|S_i|\ \ \biggr |\ \biggl (S_i\in S_{\mathcal{P}}\biggr ) 
              \mbox{ and } \biggl (S_i\not\subset\bigcup\limits_{S_j\in E_x} S_j\biggr )\biggr\};$$
\item[2)] for scenarios $w\!\in\!\mbf{w}_{\mathcal{P}}$, obtain the possible variants of $E_x$:\\ $\mbox{variant 1:} \ E_x\leftarrow E_x\cup \{S_{i_1}\},\ldots, \mbox{variant $t$:} \ E_x\leftarrow E_x\cup \{S_{i_t}\}$;  
\item[3)] obtain the SCP instances $\mathcal{P}^{(i_1)},\ldots, \mathcal{P}^{(i_t)}$ with the sets $U_{\mathcal{P}^{(i_j)}}$, $S_{\mathcal{P}^{(i_j)}}$, $\mbf{w}_{\mathcal{P}^{(i_j)}}$, $j=\overline{1,t}$.

\end{itemize}

}}

\bigskip

\noindent The set $Q$ is a collection of such sets $S_j\!\in\!S_{\mathcal{P}}$ that there is a scenario in $\mbf{w}$ for which the selection function' value on the set $S_j$ is minimal. On every iteration of the interval greedy algorithm, having an SCP instance $\mathcal{P}$ and the set $Q$, we obtain a collection of SCP instances $\mathcal{P}^{(i_1)}, \ldots, \mathcal{P}^{(i_t)}$. For all of these instances, we perform the iterations of the presented form.
\setlength{\unitlength}{0.05cm}
\begin{center}
\begin{picture}(300,120)

\put(127,109){$E_x\leftarrow\varnothing,$}
\put(127,99){$\mathcal{P}^{(0)}=\mathcal{P}$}
\put(106,117){\line(1,0){75}}
\put(106,95){\line(1,0){75}}
\put(106,95){\line(0,1){22}}
\put(181,95){\line(0,1){22}}

\put(51,68){$E_x\leftarrow E_x\cup\{ S_{i_1}\},$}
\put(80,58){$\mathcal{P}^{(i_1)}$}
\put(46,77){\line(1,0){75}}
\put(46,55){\line(1,0){75}}
\put(46,55){\line(0,1){22}}
\put(121,55){\line(0,1){22}}

\put(180,68){$E_x\leftarrow E_x\cup\{ S_{i_q}\},$}
\put(210,58){$\mathcal{P}^{(i_q)}$}
\put(176,77){\line(1,0){75}}
\put(176,55){\line(1,0){75}}
\put(176,55){\line(0,1){22}}
\put(251,55){\line(0,1){22}}

\put(5,28){$E_x\leftarrow E_x\cup\{ S_{j_1}\},$}
\put(25,18){$\mathcal{P}^{(i_1,j_1)}$}
\put(0,37){\line(1,0){75}}
\put(0,15){\line(1,0){75}}
\put(0,15){\line(0,1){22}}
\put(75,15){\line(0,1){22}}

\put(97,28){$E_x\leftarrow E_x\cup\{ S_{j_q}\},$}
\put(120,18){$\mathcal{P}^{(i_1,j_q)}$}
\put(92,37){\line(1,0){75}}
\put(92,15){\line(1,0){75}}
\put(92,15){\line(0,1){22}}
\put(167,15){\line(0,1){22}}

\put(145,95){\vector(-4,-1){67}}
\put(145,95){\vector(4,-1){67}}

\put(85,55){\vector(-4,-1){67}}
\put(85,55){\vector(4,-1){67}}

\put(216,55){\vector(-4,-1){17}}
\put(216,55){\vector(4,-1){17}}

\put(35,15){\vector(-4,-1){17}}
\put(35,15){\vector(4,-1){17}}

\put(130,15){\vector(-4,-1){17}}
\put(130,15){\vector(4,-1){17}}

\put(143,62){$\ldots$}
\put(212,42){$\ldots$}
\put(80,20){$\ldots$}
\put(31,5){$\ldots$}
\put(126,5){$\ldots$}

\put(90,-10){$\mbox{\bf{Fig. 1}. Tree of SCP instances}$}

\end{picture}
\end{center}

\bigskip

Let us give a detailed description of the procedures that the interval greedy algorithm for SCP uses. 

The procedure $\proc{Selection}$ has an SCP instance $\mathcal{P}$ as an input and it gives the set $Q$ as an output. $S_i\!\in\! Q$ only if there is a such scenario $w\!\in\!\mbf{w}_{\mathcal{P}}$ that the set $S_i$ has a minimum relative weight among the sets in $S_{\mathcal{P}}$. Let $\mbf{v}_i$ denotes the interval of {\it relative weights} of the set $S_i$: $\mbf{v}_i=\mbf{w}_i/|S_i|$.

\begin{codebox}
\Procname{$\proc{Selection}\ (\mathcal{P}): Q;$}
\li \For $\forall S_i\in S_{\mathcal{P}}$:
		   \Do
\li		  	$\mbf{v}_i\leftarrow \mbf{w}_i/|S_i|$;
			 \End	
\li $Q\leftarrow \varnothing$;
\li $v\leftarrow \min\{\overline{\mbf{v}}_i\ |\ (S_i\in S_{\mathcal{P}})\ \mbox{and}\ (S_i\neq\varnothing) \}$;
\li \For $\forall S_i\in S_{\mathcal{P}}$:
 		\Do
\li 	\If $\underline{\mbf{v}}_i\le v$	
\li 	\Then 
         $Q\leftarrow Q\cup\{S_i\}$;
			\End
		\End
		\End
\li output $Q$.		
\end{codebox}

\noindent For example, let $S\!=\!\{S_1,S_2,S_3\}$ and the intervals $\mbf{v}_i$ are $\mbf{v}_1\!=[1,5]$, $\mbf{v}_2\!=[3,7]$ and $\mbf{v}_3\!=[6,11]$ (Fig. 2). In the course of operation of the procedure $\proc{Selection}$, we select the sets $S_1$ and $S_2$ into $Q$, while the set $S_3$ we do not select into $Q$.

\setcounter{figure}{1}    

\begin{figure}[htbp]
\centering
\includegraphics[width=45mm]{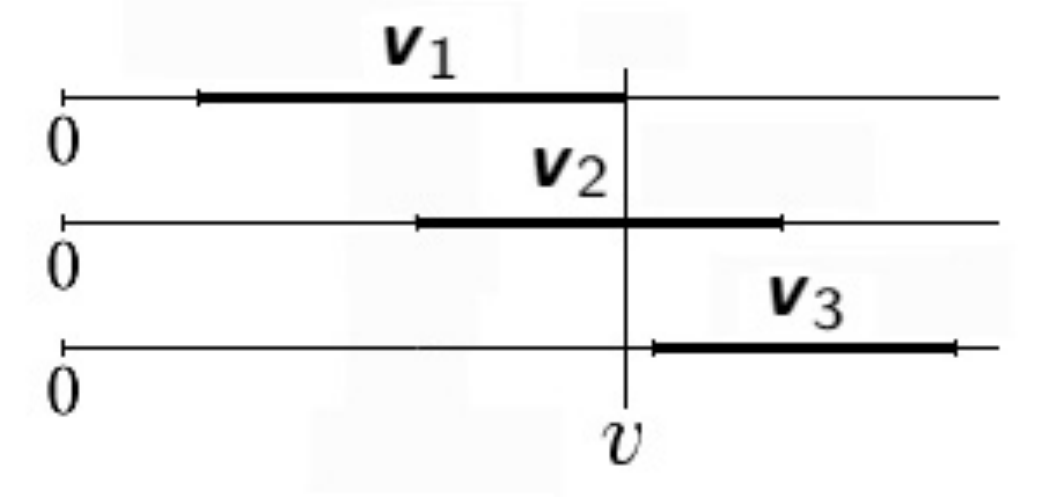}
\caption{Interval relative weights for an SCP instance}
\end{figure}

The procedure $\proc{Possible weights of a selected set}$ takes as an input an SCP instance $\mathcal{P}$ and the index $q$ of a some set that belongs to $Q$. As a result of its implementation, we have the modified interval weight $\mbf{w}_q[x]$ of the set. We obtain $\mbf{w}_q$ excluding from it the weights which are incompatible with selection of $S_q$ by the greedy algorithm for $w\!\in\!\mbf{w}[x]$.

\begin{codebox}
\Procname{$\proc{Possible weights of a selected set}\ (\mathcal{P},Q,q):\mbf{w}_q[x];$}
\li $v\leftarrow \min\{\overline{\mbf{w}}_i/|S_i|\ |\ (S_i\in Q)\ \mbox{and}\ (i\neq q)\}$;
\li \If $\overline{\mbf{w}}_{q}/|S_{q}|>v	$
\li \Then
    $\overline{\mbf{w}}_{q}\leftarrow |S_{q}|\cdot v$.
    \End
\li $\mbf{w}_q[x]\leftarrow \mbf{w}_{q}$;
\li output $\mbf{w}_q[x]$.
\end{codebox}

\noindent For the relative weights $\mbf{v}_1\!=[1,5]$, $\mbf{v}_2\!=[3,7]$ and $\mbf{v}_3\!=[6,11]$, if we take $S_2$ into $E_x$, we do not include into $\mbf{w}_2[x]$ the part of $\mbf{w}_2$ that contains the values of $w_2$ which are greater than $|S_2|\cdot v$, i.e., we exclude the values $w_2\!\in\!\mbf{w}_2$ for which relative weights $v_2\!=\!w_2/|S_2|$ are greater than $\overline{v}_1$. The set $S_2$ will not be taken by the non-interval greedy algorithm when $w\!\in\!\mbf{w}[x]$.

The procedure $\proc{Modification of an SCP instance}$ takes an SCP instance $\mathcal{P}$ and the index $q$ of the set $S_q\!\in\! Q$ as an input. For the procedure's output $\mathcal{P}'$, we have $U_{\mathcal{P}'}\!=\!U_{\mathcal{P}}\!\setminus\! S_{q}$. For the sets $S_i\not\subset\bigcup_{S_j\in E_x} S_j$, we put $S_i'=S_i\!\setminus\! S_{q}$. Also, we modify interval weights of the sets $S_i\!\in\! S_{\mathcal{P}}$ excluding the weights that are incompatible with selection of $S_{q}$ into $E_x$ by non-interval greedy algorithm for $\mbf{w}[x]$. 

\begin{codebox}
\Procname{$\proc{Modification of an SCP instance}\ (\mathcal{P},q): \mathcal{P}';$}
\li $S_{\mathcal{P}'}\leftarrow \varnothing$;
\li \For $\forall S_i\in S$ 
      \Do
\li		\If $(i\neq q)\ \mbox{and}\ (S_i\neq\varnothing)$
        \Then 
\li				\If $\underline{\mbf{w}}_i/|S_i|<\underline{\mbf{w}}_{q}/|S_{q}|$
\li				\Then 
						$\underline{\mbf{w}}'_i\leftarrow |S_i|\cdot\underline{\mbf{w}}_{q}/|S_{q}|$, $\overline{\mbf{w}}'_i\leftarrow\overline{\mbf{w}}_i$;
\li					\Else 
     				$\mbf{w}'_i\leftarrow \mbf{w}_i$;					
					\End
\li				$S_i'\leftarrow S_i\setminus S_{q}$;
			  \End
\li	   $S_{\mathcal{P}'}\leftarrow S_{\mathcal{P}'}\cup\{S_i'\}$;			  
			 \End 
		 \End
\li $U_{\mathcal{P}'}\leftarrow U_{\mathcal{P}}\setminus S_{q}$;
\li $\mbf{w}_{\mathcal{P}'}\leftarrow (\mbf{w}_1',\ldots,\mbf{w}_n')$;
\li output $\mathcal{P}'$.
\end{codebox}

\noindent For the situation that presented on Fig. $2$, taking the set $S_2$ into $E_x$, we exclude from $\mbf{w}_1$ the values $w_1$ for which the relative weights are less than $\underline{\mbf{v}}_2$. 

Note that some sets in $S_{\mathcal{P}'}$ may become empty at some iteration. To have the same enumeration for the sets in the course of the algorithm's operation, these sets are not excluded from $S_{\mathcal{P}'}$ in such situations.

The interval greedy algorithm is implemented by the following procedure.

\begin{codebox}
\Procname{$\proc{The interval greedy algorithm for SCP}\ (\mathcal{P}):\widetilde{\varXi};$}
\li $\widetilde{\varXi}\leftarrow\varnothing$; $x\leftarrow \varnothing$;
\li $\widetilde{\varXi}\leftarrow\proc{United approximate solution}\ (\mathcal{P}, x, \widetilde{\varXi});$
\li output $\widetilde{\varXi}$.
\end{codebox}

\noindent Here, the procedure $\proc{United approximate solution set}$ is the presented below backtracking procedure that use the procedures which were presented above. Implementing the procedure for an SCP instance $\mathcal{P}$ with interval weights, we obtain a weak approximate solutions which we include into united approximate solution $\widetilde{\varXi}$. The current $x$~and $\widetilde{\varXi}$ are the procedure's arguments which  are alterable during the procedure operation.

\begin{codebox}
\Procname{$\proc{United approximate solution set}\ (\mathcal{P}, x, \widetilde{\varXi});$}
\li \If $U=\varnothing$
		\Then
\li     save the pair $(E_x,\mbf{w}[x])$;
\li     $\widetilde{\varXi}\leftarrow\widetilde{\varXi}\cup\{x\}$;
\li     {\bf return}.		
\li \Else
\li			$Q\leftarrow\proc{Select}\ (\mathcal{P})$;
\li			\For $\forall S_i\in Q$:
					 \Do
\li					  $x'\leftarrow x$, $\mbf{w}'\leftarrow\mbf{w}[x]$;
\li					 	$E_x\leftarrow E_x\cup \{S_i\}$;
\li           $\mbf{w}_i[x]\leftarrow\proc{Possible weights of a selected set}\ (\mathcal{P},Q,i);$
\li						$\mathcal{P}'\leftarrow\proc{Modification of an SCP instance}\ (\mathcal{P},i)$;
\li						$\proc{United approximate solution}\ (\mathcal{P}', x, \widetilde{\varXi});$
\li					  $x\leftarrow x'$; $\mbf{w}[x]\leftarrow\mbf{w}'$. 
					 \End
		\End
\end{codebox}

Evidently, the following proposition is true.

\smallskip

\noindent{\bf Proposition.} {\it Let } $x\!=\!\{S_{i_1},\ldots,S_{i_k}\}\!\in\!\widetilde{\varXi}$. {\it Then } $$\underline{\mbf{w}}(x)=\sum_{j=1}^k\underline{\mbf{w}}_{i_j}[x],\ \overline{\mbf{w}}(x)=\sum_{j=1}^k\overline{\mbf{w}}_{i_j}[x].$$

\smallskip

As a result of the algorithm's implementation, we obtain the odered collection of sets $E_{x}$. In order to use unordered covers, which is natural for aplications, we must unite the sets of scenarios $\mbf{w}[x]$ which are obtained for different ordered $E_x$. Note that the union of the sets $\mbf{w}[x]$ may be disjoint.

The presented algorithm is a generalization of the greedy algorithm for interval weights. If all of the intervals' weights are degenerated, i.e., $\overline{\mbf{w}}_i=\underline{\mbf{w}}_i$ for all $S_i\!\in\! S$, the interval greedy algorithm operates like the non-interval greedy algorithm except the fact that it searches for not one but all possible greedy solutions if the minimum value at the step 4 of $\proc{Selection}$ is shared by several sets $S_i\!\in\!S_{\mathcal{P}}$.

\subsection{Accuracy of the solutions in $\widetilde{\varXi}$ for SCP}

SCP is $NP$-hard. The complexity of the greedy algorithm for SCP with real-valued weights is equal to $O(m^2n)$. For the general case of the problem, it holds \cite{Chvatal} that 
\begin{equation}\label{3}
w(x)\le H(m)w(\dot{x})\le (\ln m+1)w(\dot{x}),
\end{equation} 
where  $H(m)\!=\!\sum_{k=1}^m1/k$, $x$ is a cover that is obtained by the greedy algorithm, $\dot{x}$ is an optimal cover. It is shown that, whenever $\mbox{NP}\!\not\subseteq\!\mbox{TIME}(n^{O(\log\log n)})$, there is no polynomial algorithm for SCP with approximation ratio $(1-\varepsilon)\ln m$ for $\varepsilon>0$ \cite{Feige}. There are other inapproximability results for SCP which exclude the possibility of a polynomial time approximation with better than logarithmic approximation ratio.

\subsection{Computation of the weak solutions' probabilities}

In addition to intervals of possible values of element's weights, we may have some probability distribution on the intervals. Suppose a uniform probability distribution is given for the values of weights $w_i$ on the intervals $\mbf{w}_i$. This implies the uniform distribution on the intervals of relative weights. The uniform distribution is the least informative distribution of all possible distributions which is formally proved in \cite{KreinovichShary}. Further, we generalize the obtained below formulas for arbitrary probability distribution on the intervals $\mbf{w}_i$ and for an arbitrary selection function. 

The {\it probability of the weak approximate solution} $\mbox{\sffamily P}(x)$ is the probability of obtaining of a such scenario $w\!\in\!\mbf{w}$ that the non-interval greedy algorithm gives $x$ (or the set $E_x$). For the ordered set $E_{x}\!=\!\{e_{i_1},\ldots,e_{i_k}\}$, the probability $\mbox{\sffamily P}(x)$ may be computed as $\mbox{\sffamily P}(x)=\mbox{\sffamily P}(e_{i_1})\cdot\ldots\cdot\mbox{\sffamily P}(e_{i_k})$, where $\mbox{\sffamily P}(e_{i_j})$ is the probability that we take $e_{i_j}$ into $E_x$ as the $j$-th set in it performing the greedy algorithm on $w\!\in\!\mbf{w}[x]$ for current $\mbf{w}[x]$. 

The procedure $\proc{Probability of selection}$ that we shall introduce takes an SCP instance $\mathcal{P}$ and the index $q$ of $e_q\!\in\! Q$ for which we compute the probability of obtaining such $v_q\!\in\!\mbf{v}_q$ that we take $e_q$ into $E_x$ by the greedy algorithm for $w\!\in\!\mbf{w}[x]$.

To compute $\mbox{\sffamily P}(e_q)$ it uses the procedures $\proc{Partition}$ and $\proc{Probability}$. Implementing the procedure $\proc{Partition}$, we get the partition $P$ of the weights' intervals for the elements that belong to $Q$. We shall use it further in procedure $\proc{Probability}$.

\begin{codebox}
\Procname{$\proc{Partition}\ (\mathcal{P}, Q): P;$}
\li \For $\forall S_i\in Q$:
		   \Do
\li		  	$\mbf{v}_i\leftarrow\mbf{w}_i/|S_i|$;
			 \End	
\li $v_r\leftarrow\min\{\overline{\mbf{v}}_i\ |\ e_i\in Q\}$;
\li \For $\forall e_i\in Q$:
			 \Do	
\li		  	$\overline{\mbf{v}}_i\leftarrow v_r$;
			 \End	
\li $M\leftarrow \{v_1,\ldots,v_l\}$, where $v_j$ such that $\exists e_i\!\in\!Q$ for which $v_j\!=\!\underline{\mbf{v}}_i$ or $v_j\!=\!\overline{\mbf{v}}_i$,\\ $v_l\!=\!v_r$. $M$ is an ordered set and its elements are sorted in ascending order.				
\li \For $\forall e_i\in Q$:
   		 \Do
\li		 		$M_i\leftarrow \{v_j\!\in\! M\ |\ v_j\!\in\!\mbf{v}_i\}$,\\
					\quad\quad $M_i$ is an ordered set and its elements are sorted in ascending order.				   		 		
\li			$P_i\leftarrow \{\mbf{v}_{i1},\ldots,\mbf{v}_{il}\}$, where
				$\mbf{v}_{ik}\!=\![v_j,v_{j+1}]$, 
				$v_j, v_{j+1}\!\in\! M_i$.		
   		 \End 
\li $P\leftarrow \{P_1,\ldots,P_{|Q|}\}$.		
\end{codebox}

\begin{figure}[htbp]
 \centering
\includegraphics[width=45mm]{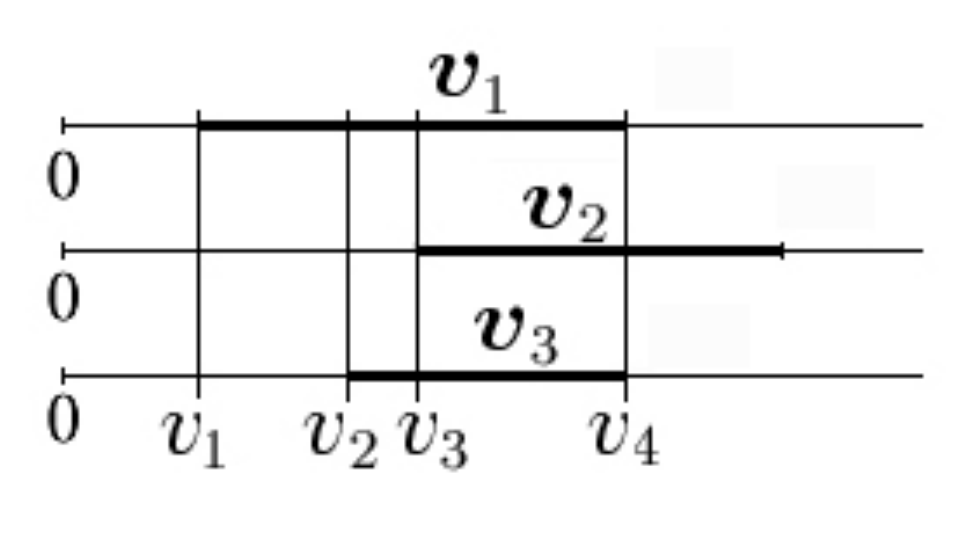}
\caption{Construction of the partition of }
\end{figure}

\noindent The Fig. 3 demonstrates construction of the partition $P$ for  $Q$, $|Q|\!=\!3$. Here we have three intervals of relative weights for three sets in $Q$. They are $\mbf{v}_1$, $\mbf{v}_2$, $\mbf{v}_3$. In this example $v_r\!\!=\!\!\overline{\mbf{v}}_1\!\!=\!\!v_4$. Constructing the partition $P\!=\!\{P_1,P_2,P_3\}$ we exclude the part of $\mbf{v}_2$ which lies on the right of $v_4$ since it contains such values of $v_2$ which are incompatible with selection of $S_2$ by the non-interval greedy algorithm. Thus we have $P_1\!=\!\{\mbf{v}_{11},\mbf{v}_{12},\mbf{v}_{13}\}$, where $\mbf{v}_{11}\!=\![v_1,v_2]$, $\mbf{v}_{12}\!=\![v_2,v_3]$, $\mbf{v}_{13}\!=\![v_3,v_4]$; $P_2\!=\!\{\mbf{v}_{21}\}$, $\mbf{v}_{21}\!=\![v_3,v_4]$; $P_3\!=\!\{\mbf{v}_{31},\mbf{v}_{32}\}$, $\mbf{v}_{31}\!=\![v_2,v_3]$, $\mbf{v}_{32}\!=\![v_3,v_4]$.

Having the partition $P$, using the procedure $\proc{Probability}$, we compute the probability $\mbox{\sffamily P}(e_q)$ of inclusion of $e_q$ into $E_x$. For a random weight $v_q$ of $e_q$, we have 
$$\mbox{\sffamily P}(e_q)=\sum_{\mbf{v}_{qj}\in P_{q}}
\mbox{\sffamily P}(e_q\ |\ v_q\in\mbf{v}_{qj})\cdot\mbox{\sffamily P}(v_{q}
\in\mbf{v}_{qj}).$$ Here $\mbox{\sffamily P}(e_q\ |\ v_q\!\in\!\mbf{v}_{qj})$ is the probability of selection of $e_q$ into cover by the greedy algorithm for $w\!\in\!\mbf{w}[x]$ for the case when $v_q\!\in\!\mbf{v}_{qj}$. For the uniform distribution on $\mbf{v}_q$, we have 
$$\mbox{\sffamily P}(v_{q}\in\mbf{v}_{qj})=(\overline{\mbf{v}}_{qj}-\underline{\mbf{v}}_{qj})/(\overline{\mbf{v}}_{q}-\underline{\mbf{v}}_{q}).$$ 

Let $$R=\{i\ |\ (i\neq q)\mbox{ and }(e_i\in Q)\mbox{ and }(\mbf{v}_{qj}\!\in\! P_i)\},$$ i.e., $R$ is a set of indices of the elements that belong to $Q\setminus\{q\}$ and the partition of its interval weights contains the interval $\mbf{v}_{qj}$. We compute the probability $\mbox{\sffamily P}(e_q\ |\ v_{q}\!\in\!\mbf{v}_{qj})$ using the formula
\begin{equation}\label{4}
\mbox{\sffamily P}(S_{q}\ |\ v_{q}\in\mbf{v}_{qj})=
\sum\limits_{r\in R}\frac{1}{|r|+1} \mbox{\sffamily P}_r,
\end{equation}
where $r$ is a subset of $R$, $\mbox{\sffamily P}_r$ is the probability of the selection of $e_q$ for a current $r$. We have $v_i\!\in\!\mbf{v}_{qj}$ for $i\!\in\!r$ and $v_i\!>\!\overline{\mbf{v}}_{qj}$ for $i\!\not\in\! r$. Denoting as $\mbox{\sffamily P}_{ij}$ probability of the event $\{v_i\!\in\!\mbf{v}_{qj}\}$ and denoting as $\mbox{\sffamily Q}_{ij}$ the probability of the event $\{v_i\!>\!\overline{\mbf{v}}_{qj}\}$, we have: $$\mbox{\sffamily P}_{ij}=(\overline{\mbf{v}}_{qj}-\underline{\mbf{v}}_{qj})/(\overline{\mbf{v}}_i-\underline{\mbf{v}}_i), \mbox{\sffamily Q}_{ij}=(\overline{\mbf{v}}_i-\overline{\mbf{v}}_{qj})/(\overline{\mbf{v}}_i-\underline{\mbf{v}}_i),$$ and thus
\begin{equation}\label{5}
\mbox{\sffamily P}_r=\prod_{i\in r}\mbox{\sffamily P}_{ij}\prod_{i\in R\setminus r}\mbox{\sffamily Q}_{ij},\end{equation}
where $r$ takes all possible values in (\ref{5}) ranging from $r\!=\!\varnothing$ to $r\!=\!R$. If $r\!=\!\varnothing$ or $R\setminus r\!=\!\varnothing$, we substitiute the corresponding product ($\prod_{i\in r}\mbox{\sffamily P}_{ij}$ or $\prod_{i\in R\setminus r}\mbox{\sffamily Q}_{ij}$) by $1$ in (\ref{5}). The multiplier $1/(1+|r|)$ is due to the fact that, for random variables $\xi_1,\ldots,\xi_N$ which are uniformly distributed on the same interval, we have $\mbox{\sffamily P}(\xi_{i}\!=\!\min\{\xi_1,\ldots,\xi_N\})\!=\!1/N$ for all $i$. 

To compute the probability $\mbox{\sffamily P}(e_q)$, we implement the procedure $\proc{Probability}$ $\proc{of selection}$.

\begin{codebox}
\Procname{$\proc{Probability of selection}\ (\mathcal{P},Q,q): \mbox{\sffamily P}(e_q);$}
\li $P\leftarrow \proc{Partition}\ (\mathcal{P}, Q)$;
\li $\mbox{\sffamily P}(e_q)\leftarrow \proc{Probability}\ (\mathcal{P},P,Q,q)$.
\end{codebox}

\begin{codebox}
\Procname{$\proc{Probability}\ (\mathcal{P},P,Q,q): \mbox{\sffamily P}(e_q);$}
\li \For $j \gets 1$ \To $|P_q|$ 
       \Do
\li $\mbox{\sffamily P}(v_{q}\in\mbf{v}_{qj})=(\overline{\mbf{v}}_{qj}-\underline{\mbf{v}}_{qj})/(\overline{\mbf{v}}_{q}-\underline{\mbf{v}}_{q})$;
\li       $R\leftarrow \{i\in Q\ |\ \mbf{v}_{qj}\in P_i\}$;
\li    		\For $\forall i\in R$:
       				\Do
\li							$\mbox{\sffamily P}_{ij}\leftarrow (\overline{\mbf{v}}_{qj}-\underline{\mbf{v}}_{qj})/
								(\overline{\mbf{v}}_i-\underline{\mbf{v}}_i)$;
       				\End
\li    		\For $\forall i\in R\setminus r$:
       				\Do
\li							$\mbox{\sffamily Q}_{ij}\leftarrow (\overline{\mbf{v}}_i-\overline{\mbf{v}}_{qj})/
								(\overline{\mbf{v}}_i-\underline{\mbf{v}}_i)$;       				
							\End

\li			 

$\mbox{\sffamily P}(e_q\ |\ v_{q}\in\mbf{v}_{qj})\leftarrow \sum\limits_{r\in R}\biggl (\dt\frac{1}{|r|+1}\prod\limits_{i\in r}\mbox{\sffamily P}_{ij}\prod\limits_{i\in R\setminus r}\mbox{\sffamily Q}_{ij}\biggr )$;	       
       \End
		
\li $\mbox{\sffamily P}(e_q)\leftarrow \sum\limits_{j=1}^{|P_q|}\mbox{\sffamily P}(e_q\ |\ v_{q}\in\mbf{v}_{qj})\cdot\mbox{\sffamily P}(v_q\in\mbf{v}_{qj})$.
\end{codebox}

If there are such sets $S_q$, $q\!\in\! Q$, that $\mbf{v}_q$ is degenerated, we do the computation of $\mbox{\sffamily P}(e_q)$ using the formula  
\begin{equation}\label{6}
\mbox{\sffamily P}(e_q)=\prod\limits_{i\in Q, i\neq q}[(\overline{\mbf{v}}_i-v_q)/(\overline{\mbf{v}}_i-\underline{\mbf{v}}_i)].
\end{equation} For $i\!\in\! Q$, $i\!\neq\! q$, such that $\mbf{v}_i$ is degenerated and $\mbf{v}_i\!=\!\mbf{v}_q$, we replace the factor $(\overline{\mbf{v}}_i-v_q)/(\overline{\mbf{v}}_i-\underline{\mbf{v}}_i)$ in (\ref{6}) by 1.

To compute $\mbox{\sffamily P}(x)$ for $x\!\in\!\widetilde{\varXi}$, we need to do the following modifications of the interval greedy algorithm's procedures. At the step $1$ of procedure $\proc{The interval}$ $\proc{greedy algorithm}$, we set $\mbox{\sffamily P}(x)\!\leftarrow\! 1$. The procedure $\proc{Probability of a selec-}$ $\proc{tion}$ is called before implementation of the procedure $\proc{Modification of the}$ $\proc{problem instance}$ at the course of operation of the procedure $\proc{United solu-}$ $\proc{tion set}$. Taking $e_{i}$ in $E_x$, before the step $10$ of the procedure, we compute the current value $\mbox{\sffamily P}(x)$: $$\mbox{\sffamily P}(x)\leftarrow\mbox{\sffamily P}(x)\cdot\proc{Probability of selection}\ (\mathcal{P},Q,i).$$ 

The value $\mbox{\sffamily P}(x)$, that we compute before selection of $e_i$ into $E_x$, must be saved at the step $8$ of the procedure $\proc{United approximate solution set}$. It must be restored  on the step $14$ in order to compute the probabilities of other approximate solutions that we obtain taking other sets from $Q$ into $E_x$ at the iteration. The finally computed value $\mbox{\sffamily P}(x)$ must be saved before we quit the procedure on the step $4$.

\smallskip

For the $x$ that we have on an iteration of the greedy algorithm, the curent value $\mbox{\sffamily P}(x)$ is an upper bound on probability of the weak approximate solution that we may obtain for the current $x$. So, comparing $\mbox{\sffamily P}(x)$ with some given threshold value $\delta$, and not dealing with such $x\!\in\!\tilde{\varXi}$ that $\mbox{\sffamily P}(x)\!<\!\delta$ at the course of the algorithm's implementation, we may obtain only such solutions $x\!\in\!\tilde{\varXi}$ that $\mbox{\sffamily P}(x)\!\ge\!\delta$. In order to do so, we do not call the backtrack procedure $\proc{United approximate solution set}$ when $\mbox{\sffamily P}(x)\!<\!\delta$. It is clear, that such a modification of the algorithm decreases its computational complexity.

\subsection{The probability distribution\\ on the set of posible values of objective function}

Suppose that, in the course of the greedy algorithm operating, we get {\it the set of possible objective function's values} $\mbf{\mathrm{w}}(\widetilde{\varXi})$ for united approximate solutions set $\widetilde{\varXi}$. We have 
$$
\mbf{\mathrm{w}}(\widetilde{\varXi})=\bigcup\limits_{x\in\widetilde{\varXi}}\mbf{w}(x)=\bigcup_{x\in\widetilde{\varXi}}\biggl (\sum\limits_{S_i\in E_x}\mbf{w}_i[x]\biggr ).
$$ The set $\mbf{\mathrm{w}}(\widetilde{\varXi})$ may be a collection of disjoint intervals.

Having the probabilities $\mbox{\sffamily P}(x)$ for $x\!\in\!\widetilde{\varXi}$, we may compute the probability distribution on $\mbf{\mathrm{w}}(\widetilde{\varXi})$. Let $\mathrm{w}\!\in\!\mbf{\mathrm{w}}(\widetilde{\varXi})$ be the possible weight of a solution in $\widetilde{\varXi}$. The value of $\mathrm{w}$  depends on the random scenario $w\!\in\!\mbf{w}$. Let $\mbox{\sffamily p}(\mathrm{w}|x)$ be the density of $\mathrm{w}$ on the interval weight $\mbf{w}(x)$. Then, for the density $\mbox{\sffamily p}(\mathrm{w})$ of the distribution of $\mathrm{w}$ on $\mbf{\mathrm{w}}(\widetilde{\varXi})$, we have 
\begin{equation}\label{10}
\mbox{\sffamily p}(\mathrm{w})\!=\!\sum\limits_{x\in\widetilde{\varXi}}\mbox{\sffamily p}(\mathrm{w}|x)\mbox{\sffamily P}(x).
\end{equation}

For $x\!\in\!\widetilde{\varXi}$ with a large enough amount of non-zero components, the density $\mbox{\sffamily p}(\mathrm{w}|x)$ tends to the density of the normal distribution for independent random values of weights as the Lindeberg (central limit) theorem states. For $x\!\in\!\widetilde{\varXi}$ with a small  amount of non-zero components, we may use convolutional formulae to compute $\mbox{\sffamily p}(\mathrm{w}|x)$ in (\ref{10}).

Having probability distribution on $\mbf{\mathrm{w}}(\widetilde{\varXi})$, we may compute mean value of the approximate solutions' weights for the problem, its dispersion and other probabilistic  characteristics of the objective function' value.

\subsection{The interval greedy algorithm\\ for discrete optimization problems}

Let us formulate the general scheme of the interval greedy algorithm for an arbitrary discrete optimization problem of the form (III). For such problems, the formulations of the non-interval greedy algorithm differ only by its selection function. Let $\varphi(e_i)$ be the real-valued selection function that used to select the elements of $E$ performing the non-interval greedy algorithm to solve the problem of the form (I). For the problem with interval weights $\mbf{w}_i$ of elements $e_i\!\in\!E$, we have interval selection function $\mbf{\varphi}(e_i)\!=\![\overline{\mbf{\varphi}}(e_i), \underline{\mbf{\varphi}}(e_i)]$. Here, $\mbf{\varphi}(e_i)$ is an interval of possible values of $\varphi(e_i)$ for scenarios in $\mbf{w}$.

All of the procedures, which are introduced below, have the same structure and justification as the procedures of the interval greedy algorithm for SCP have. All of the considerations on intervals of possible values of weights of the solutions that belong to united solution set or united aproximate solution set, the considerations on computaions of probabilities of the solutions for the case of a uniform distribution, the considerations on probability distribution on the set of possible values of objective function that we made above are valid for the general scheme of the interval greedy algorithm.

\begin{codebox}
\Procname{$\proc{Selection}\ (\mathcal{P}): Q;$}
\li $Q\leftarrow \varnothing$; $\varphi_{\min}\leftarrow \min\{\overline{\mbf{\varphi}}(e_i)\ |\ e_i\in E\setminus E_x \}$;
\li \For $\forall e_i\in E$:
 		\Do
\li 	\If $\underline{\mbf{\varphi}}(e_i)\le \varphi_{\min}$	
\li 	\Then 
         $Q\leftarrow Q\cup\{e_i\}$;
			\End
		\End
		\End
\li output $Q$.		
\end{codebox}

\begin{codebox}
\Procname{$\proc{Possible weights of a selected element}\ (\mathcal{P},Q,q):\mbf{w}_q[x];$}
\li $\varphi_{\min}\leftarrow \min\{\overline{\mbf{\varphi}}(e_i)\ |\ e_i\in E\setminus E_x\}$;
\li \If $\overline{\mbf{\varphi}}(e_q)>\varphi_{\min}$
\li \Then
get $\mbf{w}_q[x]$ excluding from $\mbf{w}_q$ such $w_q$ that $\varphi(e_q)\!>\!\varphi_{\min}$;
    \End

\li output $\mbf{w}_q[x]$.
\end{codebox}

\begin{codebox}
\Procname{$\proc{Modification of the problem instance}\ (\mathcal{P},q): \mathcal{P}';$}
\li $E'\leftarrow \varnothing$;
\li \For $\forall e_i\in E$ 
      \Do
\li		  \If $i\neq q$
        \Then 
\li				\If $\underline{\mbf{\varphi}}(e_i)<\underline{\mbf{\varphi}}(e_q)$
\li				\Then 
            get $\mbf{w}'_i$ excluding from $\mbf{w}_i$ such $w_i$ that $\varphi(e_q)\!>\!\varphi(e_i)$;
\li				 \Else 
     				$\mbf{w}'_i\leftarrow \mbf{w}_i$;					
					\End
			  \End
\li	   $E'\leftarrow E'\cup\{e_i\}$;			  
			 \End 
		 \End
\li $\mbf{w}_{\mathcal{P}'}\leftarrow (\mbf{w}_1',\ldots,\mbf{w}_n')$;
\li output $\mathcal{P}'$. 
\end{codebox}

\begin{codebox}
\Procname{$\proc{The interval greedy algorithm}\ (\mathcal{P}):\widetilde{\varXi};$}
\li $E_x\leftarrow \varnothing$; $\widetilde{\varXi}\leftarrow\varnothing$; $x\leftarrow (0,\dots,0);$
\li $\widetilde{\varXi}\leftarrow \proc{United solution set}\ (\mathcal{P}, x, \widetilde{\varXi});$
\li output $\widetilde{\varXi}$.
\end{codebox}

\begin{codebox}
\Procname{$\proc{United approximate solution set}\ (\mathcal{P}, x, \varXi);$}
\li \If $x\!\in\!\mathcal{D}$
		\Then
\li     save the pair $(E_x, \mbf{w}[x])$;
\li     $\widetilde{\varXi}\leftarrow\widetilde{\varXi}\cup\{x\}$;
\li     {\bf return}.		
\li \Else
\li			$Q\leftarrow\proc{Select}\ (\mathcal{P})$;
\li			\For $\forall e_i\in Q$:
					 \Do
\li					  $x'\leftarrow x$; $\mbf{w}'\leftarrow\mbf{w}[x]$;	$E_x\leftarrow E_x\cup \{e_i\}$;
\li           $\mbf{w}_i[x]\leftarrow\proc{Possible weights of a selected element}\ (\mathcal{P},Q,i);$
\li						$\mathcal{P}'\leftarrow\proc{Modification of the problem instance}\ (\mathcal{P},i)$;
\li						$\proc{United approximate solution set}\ (\mathcal{P}', x, \widetilde{\varXi});$
\li					  $x\leftarrow x'$; $\mbf{w}[x]\leftarrow\mbf{w}'$. 
					 \End
		\End
\end{codebox}

\paragraph{Probabilities of approximate weak solutions.} For an arbitrary interval selection function $\mbf{\varphi}$, for given uniform probability distribution on its values, replacing the symbols $w$ and $\mbf{w}$ by symbols $\varphi$ and $\mbf{\varphi}$ respectively in all of the notations above, preserving its subscripts and superscripts, we obtain the formulae to compute the probabilities of the weak (approximate) solutions.

For the case of an arbitrary probability distribution, let $\mbox{\sffamily p}_i(t)$ be a probability density function for the weights distribution on interval $\mbf{\varphi}_i$. Then the probability that $\varphi_i\!\in\![a,b]$ is $$\mbox{\sffamily P}_{[a,b]}(\varphi_i)=\int\limits_a^b\mbox{\sffamily p}_i(t) dt.$$ So the probabilities $\mbox{\sffamily P}_{ij}$ and $\mbox{\sffamily Q}_{ij}$ are the following for the general case:
\begin{equation}\label{7}
\mbox{\sffamily P}_{ij}=\int\limits_{\underline{\mbf{\varphi}}_{qj}}^{\overline{\mbf{\varphi}}_{qj}}\mbox{\sffamily p}_i(t) dt,\quad \mbox{\sffamily Q}_{ij}=\int\limits_{\overline{\mbf{\varphi}}_i}^{\overline{\mbf{\varphi}}_{qj}}\mbox{\sffamily p}_i(t) dt,
\end{equation}
and 
\begin{equation}\label{8}
\mbox{\sffamily P}(\varphi_{q}\in\mbf{\varphi}_{qj})=\int\limits_{\underline{\mbf{\varphi}}_{qj}}^{\overline{\mbf{\varphi}}_{qj}}\mbox{\sffamily p}_q(t) dt,
\end{equation}
where ${\mbf{\varphi}}_{ij}$ are elements of the partition that we builds on intervals $\mbf{\varphi}_i$, $i\!\in\!Q$. 

For the sets $S_q$, $q\!\in\! Q$, such that $\mbf{\varphi}_q$ is degenerated, we compute $\mbox{\sffamily P}(e_q)$ using the formula  
\begin{equation}\label{9}
\mbox{\sffamily P}(e_q)=\prod\limits_{i\in Q, i\neq q}\int\limits_{\mbf{\varphi}_q}^{\overline{\mbf{\varphi}}_i}\mbox{\sffamily p}_i(t) dt.
\end{equation}
For $i\!\in\! Q$, $i\!\neq\! q$, such that $\mbf{\varphi}_i$ is  degenerated, we replace the corresponding factor in (\ref{9}) by 1. 

Using the formulae (\ref{7})--(\ref{9}) for computing the values $\mbox{\sffamily P}(e_q)$, $\mbox{\sffamily P}_{ij}$ and $\mbox{\sffamily Q}_{ij}$, we may compute the probabilities $\mbox{\sffamily P}(x)$, $x\!\in\!\widetilde{\varXi}$, for arbitrary probability distribution that is given on intervals of weights and for arbitrary selection function. For the case of discrete probability distribution, when all of the weights are degenerated, the Stieltjes integration is applied in (\ref{7}) -- (\ref{9}).

\subsection{Computational complexity of the approach}

Complexity of the interval greedy algorithm greatly depends on the instance of the problem of the form (III) to which its applied. It is exponential at the worst case. Its complexity depends on the number of calls of the backtracking procedure $\proc{United approximate solution set}$, i.e., the complexity depends on the values of $|Q|$ that we obtain performing the procedure $\proc{Selection}$. An interval vector $\mbf{w}_{\mathcal{P}}$ and combinatorial structure of the problem instance $\mathcal{P}$ of the form (III) are determine the search tree and, consequently, they determine the computational complexity of solution of an instance $\mathcal{P}$ by the interval greedy algorithm. 

As it was shown in \cite{Prolubnikov2}, the complexity is a non-decreasing step function on values of radii of the weights' intervals for SCP. And, since the complexity of the algorithm depends only on mutual positions of the weights' intervals \cite{Prolubnikov2}, the result may be applied to the general case of discrete optimization problems of the form (III).

\section{Conclusions}

We consider a wide class of discrete optimization problems with interval objective functions. We give a generalization of the greedy algorithm for the class. Applying the presented approach to the problem $\mathcal{P}$ of the form (III), we may obtain the following information: 
\begin{itemize}
\item[1)] the united approximate solution set $\widetilde{\varXi}$; 
\item[2)] the sets of scenarios $\mbf{w}[x]\!\subseteq\!\mbf{w}_{\mathcal{P}}$ for $x\!\in\!\widetilde{\varXi}$;
\item[3)] the intervals $\mbf{w}(x)$ of possible weights for $x\!\in\!\widetilde{\varXi}$: $\mbf{w}(x)=\sum_{S_i\in E_{x}}\mbf{w}_i[x]$; 
\item[4)] the probabilities of $x\!\in\!\widetilde{\varXi}$ for a given probability distribution on weights' intervals; 
\item[5)] the probability distribution on the set of possible objective function' values $\mbf{\mathrm{w}}(\widetilde{\varXi})$ for solutions that belongs to $\widetilde{\varXi}$. Using the distribution, we may compute expected value of the objective function, its standard deviation, etc.
\end{itemize}

%
%
%
%

\end{document}